\documentclass[letterpaper, 10 pt, conference]{ieeeconf}  % Comment this line out if you need a4paper
\IEEEoverridecommandlockouts                              % This command is only needed if 
\usepackage[english]{babel}
\usepackage{cite}
\usepackage{amsmath,amssymb,amsfonts}
\usepackage{algorithmic}
\usepackage{graphicx}
\usepackage[labelformat=simple]{subcaption}

\usepackage{textcomp}
\usepackage{amsmath}
\usepackage{xcolor}
\usepackage{units}
\usepackage{booktabs}
\usepackage{comment}
\usepackage[hidelinks]{hyperref}
\usepackage[normalem]{ulem}

\usepackage{amsthm}
\usepackage{flushend}
\usepackage{setspace}

\newtheorem{theorem}{Theorem}
\newtheorem{lemma}{Lemma}

\newtheorem{definition}{Definition}
\newtheorem{remark}{Remark}

\newtheorem{problem}{Problem}

\newif\ifextendedversion
\extendedversiontrue

\usepackage{endnotes}

\newcommand{\dint}[0]{\mathrm{d}}
\DeclareMathOperator{\argmin}{\mathrm{argmin}}
\DeclareMathOperator{\EV}{\mathbb{E}}
\DeclareMathOperator{\Var}{\mathrm{Var}}

\newcommand{\PoA}[0]{\mathrm{PoA}}
\newcommand{\BRS}[0]{\mathrm{BRS}}
\newcommand{\InEqt}[0]{\mathrm{InEqt}}
\newcommand{\InEql}[0]{\mathrm{InEql}}
\DeclareMathOperator{\rat}{\mathrm{rat}}

\newcommand{\N}[0]{\mathbb{N}}
\newcommand{\Nz}[0]{\mathbb{N}_0}
\newcommand{\R}[0]{\mathbb{R}}
\newcommand{\Rp}[0]{\mathbb{R}_{>0}}
\newcommand{\Rnn}[0]{\mathbb{R}_{\geq0}}
\newcommand{\Q}[0]{\mathbb{Q}}
\newcommand{\Qp}[0]{\mathbb{Q}_{>0}}

\newcommand{\Acal}[0]{\mathcal{A}}
\newcommand{\Fcal}[0]{\mathcal{F}}
\newcommand{\Kcal}[0]{\mathcal{K}}

\newcommand{\Rcal}[0]{\mathcal{R}}
\newcommand{\Wcal}[0]{\mathcal{W}}
\newcommand{\Xcal}[0]{\mathcal{X}}

\newcommand{\Pt}[0]{\mathcal{P}_t}
\newcommand{\Ait}{\mathcal{A}^i_t}
\newcommand{\AtNE}{\mathcal{A}^{\mathrm{NE}}_t}

\newcommand{\At}[0]{A_t}
\newcommand{\nAt}[0]{\mathbf{n}^{\At}}

\newcommand{\wAt}[0]{\mathbf{w}^{\At}}
\newcommand{\wrAt}[0]{\mathbf{w}_r^{\At}}
\newcommand{\AtBRS}[0]{\At^\BRS}
\newcommand{\wAtBRS}[0]{\mathbf{w}^{\AtBRS}}

\newcommand{\Nto}[0]{N_t^{\{1\}}}

\newcommand{\wstar}[0]{\mathbf{w}^\star}
\newcommand{\wbar}[0]{\bar{\mathbf{w}}}
\newcommand{\pbar}[0]{\bar{\mathbf{p}}}
\newcommand{\pbarOj}[0]{\pbar^{\Omega_j}}
\newcommand{\nbar}[0]{\bar{\mathbf{n}}}
\newcommand{\nbarOj}[0]{\nbar^{\Omega_j}}

\newcommand{\lstar}[0]{\mathbf{l}^\star}
\newcommand{\Pgo}[0]{P_\mathrm{go}}
\newcommand{\wmin}[0]{w_\mathrm{min}}
\newcommand{\wmax}[0]{w_\mathrm{max}}

\newcommand{\colK}[0]{\{K_t\}_{t\in \Nz}}
\newcommand{\colL}[0]{\{L_t\}_{t\in \Nz}}
\newcommand{\colP}[0]{\{P_t\}_{t\in \Nz}}

\title{\LARGE \bf
Fair Artificial Currency Incentives in Repeated Weighted Congestion Games: Equity vs. Equality}

\author{Leonardo Pedroso, Andrea Agazzi, W.P.M.H. (Maurice) Heemels, Mauro Salazar% <-this % stops a space
\thanks{L. Pedroso, W.P.M.H. Heemels, and M. Salazar are with the Control Systems Technology section, Eindhoven University of Technology, The Netherlands (email: {\tt \footnotesize \{l.pedroso,m.heemels, m.r.u.salazar\}@tue.nl}). A. Agazzi is with the Mathematics Department, Università di Pisa, Pisa, Italy (email: {\tt \footnotesize andrea.agazzi@unipi.it}). A. Agazzi is member of INdAM (GNAMPA group), and acknowledges partial support of Dipartimento di Eccellenza, UNIPI, the Future of Artificial Intelligence Research (FAIR) foundation (WP2), PRIN project ConStRAINeD, PRA Project APRISE, and GNAMPA Project CUP\_E53C22001930001.}%
}

\ifextendedversion
\else
\fi

\begin{document}

\maketitle
\thispagestyle{empty}
\pagestyle{empty}

\begin{abstract}
	
	 	When users access shared resources in a selfish manner, the resulting societal cost and perceived users' cost is often higher than what would result from a centrally coordinated optimal allocation.
		While several contributions in mechanism design manage to steer the aggregate users choices to the desired optimum by using monetary tolls, such approaches bear the inherent drawback of discriminating against users with a lower income.
		More recently, incentive schemes based on artificial currencies have been studied with the goal of achieving a system-optimal resource allocation that is also fair.
		In this resource-sharing context, this paper focuses on repeated weighted congestion game with two resources, where users contribute to the congestion to different extents that are captured by individual weights. First, we address the broad concept of fairness by providing a rigorous mathematical characterization of the distinct societal metrics of equity and equality, i.e., the concepts of providing equal outcomes and equal opportunities, respectively.
		Second, we devise weight-dependent and time-invariant optimal pricing policies to maximize equity and equality, and prove convergence of the aggregate user choices to the system-optimum.
		In our framework it is always possible to achieve system-optimal allocations with perfect equity, while the maximum equality that can be reached may not be perfect, which is also shown via numerical simulations.

%		\rev{In this resource-sharing context}, this paper focuses on the broad concept of fairness by rigorously optimizing for the distinct societal metrics of equity and equality, i.e., the concepts of providing equal outcomes and equal opportunities, respectively.
%		First, we frame our problem for a repeated weighted congestion game with two resources, where users contribute to the congestion to different extents that are captured by individual weights, and provide a rigorous mathematical characterization of both metrics.

%In this paper, we address the mechanism design problem in repeated weighted congestion games from a fairness perspective. 

%Keywords for serach engines :)
% Mechanims design
% Incentive 
% Fair fairness
% Resource allocation 
% Congestion games

\end{abstract}

% !TeX spellcheck = en_US

\section{Introduction}

Recent advances in the internet of things and connectivity have led to the rise of sharing economies in today's society. In these settings, users compete for access to shared resources such as mobility infrastructures, cloud computing services, and electrical power. To harness the full potential utility of the resources, it is imperative to judiciously design rules for their allocation among the users. This is known as a mechanism design problem~\cite{ChremosMalikopoulos2024} and it has been recently flagged as an urgent societal challenge for the control community~\cite{AnnaswamyJohanssonEtAl2023}. In this paper, we consider a population of users that desires to use one resource out of a set of available resources (e.g., choosing a time slot in a day for charging an electric vehicle). Users may contribute differently to congestion (e.g., different power drawn from the grid), which is portrayed by a weight associated with each user. The discomfort perceived by a user in choosing each resource depends on the cumulative weight of users choosing that resource only. We also consider that the allocation is repeated periodically (e.g., daily). This allocation setting is called a \emph{repeated weighted congestion game} in the game-theory literature.

In these settings, each user selfishly chooses the resources that they use in each repetition in such a way that their perceived discomfort is as low as possible. It is well-known that such self-interested behavior leads to inefficient aggregate equilibria  \cite{Dubey1986}, i.e., there exists a central allocation of resources to users for which the average user discomfort could be lower.
The prime illustrative example of this inefficiency is Pigou's two-arc network \cite[Chap.~17]{NisanRoughgardenEtAl2007}. More generally, one may want to consider a societal cost that is different from the average user discomfort, for which selfish equilibria are also inefficient in general.
 Whilst a central allocation could readily solve such inefficiencies, it follows authoritarian paradigms and does not capture the users' varying temporal needs.
We may then naturally ask: Can we design a mechanism that aligns the users' self-interested behavior with a societally-optimal aggregate decision pattern? This problem has been addressed over the past 70 years and, for the problem at hand, it has been primarily addressed by levying monetary tools on each resource \cite[Chap.~18]{NisanRoughgardenEtAl2007}.

The use of monetary criteria is aligned with a profit-driven vision. However, there are scenarios with the focus on the efficient but unbiased provision of resources (e.g., in a mobility setting). In that sense, monetary criteria are unfair, since they discriminate against users with lower income. In those settings, it is then natural to consider a concept of fairness among users hand in hand with the efficiency of selfish equilibria. In a single allocation instance, efficiency and fairness are conflicting objectives, since at the societal-optimum generally some users experience more discomfort than others. Nevertheless, a repeated allocation setting allows to incentivize turn-taking behavior, whereby users take turns using the most uncomfortable resources. As a result, in a repeated setting, fairness and efficiency are no longer conflicting. A mechanism that induces the pivotal turn-taking behavior has been recently proposed \cite{CensiBolognaniEtAl2019}. It leverages an artificial currency (AC) that cannot be traded or bought for money. Each user has a wallet of AC, whenever they choose a resource they pay a corresponding price of the AC, and they are restricted to choosing resources that they can afford. The AC level of the users is, thus, a measure of how altruistic they have been in their past decisions. Several works on AC-based incentive schemes, also called Karma schemes, have followed the principle introduced in \cite{CensiBolognaniEtAl2019}, such as~\cite{SalazarPaccagnanEtAl2021,JalotaPavoneEtAl2020,ElokdaCenedeseEtAl2022,ElokdaBolognaniEtAl2023,PedrosoHeemelsEtAl2023}.

%Overall unfairness is defined as the maximum unfairness among all users

A quantification of the unfairness of a repeated allocation of resources captures the imbalance of the average perceived discomfort among the players. The first formal definition of the concept of fairness in resource allocation games was introduced in \cite{JahnMohringEtAl2005}, in a mobility setting. Therein, for a single allocation instance, the unfairness for a user is defined as the ratio between their travel time and the minimum travel time experienced by a user with the same origin-destination pair. In \cite{CensiBolognaniEtAl2019}, unfairness is defined as the variance (across the users) of the accumulated perceived discomfort of each user. However, when users have different needs, which are represented by their weights in the setting at hand, one can evaluate fairness from two perspectives: \emph{equity} and \emph{equality}. On the one hand, \emph{equity} is associated with providing the same outcome to all users, regardless of their weight. On the other hand, \emph{equality} is associated with providing the same opportunity to all users, which, in a weighted congestion game setting, means that users are given the same resource utility per unit weight. The outcome and, thus, the design for each of these perspectives may be dramatically different. For example, in a problem of power allocation to charge electric vehicles, one achieves perfect equality if the charging power is the same for every user, whereas one achieves perfect equity if the power is split in such a manner that all users achieve the same state-of-charge.
In this paper, we design two AC-based incentive schemes optimizing for each of these perspectives of fairness. We take the particular case of only two resources, as a first approach. To the best of the authors' knowledge, the equity vs. equality dichotomy in mechanism design has not been addressed previously in the literature. 

\emph{Statement of Contributions}: The main contributions of this paper are twofold. First, we propose a formal quantitative definition of equity and equality in repeated resource allocation settings. Second, for the first time in the literature, we design two optimal AC-based incentive schemes that maximize equity and equality in repeated weighted congestion games. 

\emph{Organization}: The remainder of this paper is organized as follows. In Section~\ref{sec:problem_statement}, the AC pricing problem is formally stated. In Section~\ref{sec:design}, we design the optimal AC schemes that maximize equity and equality. In Section~\ref{sec:results}, we illustrate our approach resorting to numerical simulations. Finally, in Section~\ref{sec:conclusions}, we draw the conclusions from our findings and provide an outlook on future research endeavors.

\emph{Notation}:
The vector of ones, of appropriate dimensions, is denoted by $\mathbf{1}$. The $i$th entry of a vector $\mathbf{v}$ is denoted by $\mathbf{v}_i$.  The indicator function of set $\Acal \subset  \Xcal$ is denoted by ${\mathbf{1}_\Acal :\Xcal \to \{0,1\}}$, whereby $\mathbf{1}_\Acal(x) = 1$, if $x\in \Acal$, and $\mathbf{1}_\Acal(x) = 0$, otherwise. The expected value of a random variable $X$ is denoted by $\EV[X]$.

%The cardinality of a set $\Acal$ is denoted by $|\Acal|$. 
%The Iverson bracket of a statement $p$ is denoted by $[p]$, whereby $[p] = 1$, if $p$ is true, and $[p] = 0$, otherwise. 
%The indicator function of set $\Acal \subset  \Xcal$ is denoted by ${\mathbf{1}_\Ncal :\Xcal \to \{0,1\}}$, whereby $\mathbf{1}_\Ncal(x) = 1$, if $x\in \Acal$, and $\mathbf{1}_\Ncal(x) = 0$, otherwise.

% !TeX spellcheck = en_US
\section{Problem Statement}\label{sec:problem_statement}

%It is the extension of a particular case of the overarching formulation presented in \cite{PedrosoHeemelsEtAl2024} to weighted congestion games. 

In this section, the AC pricing problem is formally stated. We consider an infinite population of players, whereby each player carries an infinitesimal amount of the population's weight. We formulate the problem relying on sequences of random variables (r.v.(s)) for a generic player given the Lebesgue measure on $\Omega \!= \![0,1]$, which is denoted by the probability triple $(\Omega,\Fcal, P)$.\footnote{By the Extension Theorem~\cite[Theorem~2.3.1]{Rosenthal2006}, the probability triple $(\Omega,\Fcal, P)$ and all r.v.s introduced in the upcoming sections are well-defined.} In this setting, a generic r.v.\ $X$ is a Lebesgue-measurable function $X\!: \!\Omega \to \R$, and the probability of an event $\Acal \!\in \!\Fcal$ (with $\Acal \subseteq \Omega$) is denoted by $P(\Acal)$. Intuitively, by abuse of terminology, each player can be regarded as a real number  $i \!\in\! \Omega \!= \![0,1]$, and $X(i)$ is the value that $X$ takes for player $i$. This approach is in line with some works on monetary mechanism design~(e.g.,~\cite{ColeDodisEtAl2003}). See \cite{Rosenthal2006} for a detailed overview of the elements of probability theory using measure theory employed in this paper.

%as well as all r.v.\ introduced in the upcoming sections are

%\notef{@M\&M: Without proper measure theory we were super constrained on the generality of the problems we could tackle (e.g., the weight distribution would have to have finite and rational support, which I do not think makes sense considering that we have an infinite population). Also, the results on the conditions for equality and equity would look super ugly. On the negative side, some readers may be scared by this, but I tried to make it as simple, clear, and intuitive as possible.}

Consider a set of two resources $\Rcal := \{1,2\}$. We consider a weighted singleton congestion game setting, that repeats at every time $t = \{0,1,\ldots\}$. The participation of the players in the game at time $t\!\in \!\Nz$ is described by independent and identically distributed (i.i.d.) r.v.s ${P_t: \Omega \to \{0,1\}}$ with Bernoulli distribution, for which $P(P_t \!=\!1) \!= \!\Pgo\! \in (0,1)$. We denote the subset of the population participating at time $t$ by $\Pt\!:=\! \{i\!\in\! \Omega : P_t(i) \!=\!1\}$. At each instance of the game, a participating player $i\in \Pt$ chooses one of the resources in $\Rcal$ to satisfy their needs, and a nonparticipating player does not choose any resource, i.e., chooses strategy $\emptyset$. The outcome of the players' decision at time $t$ is a Lebesgue-measurable function $A_t \! :\! \Omega \to \{0,1,2\}$, whereby $A_t(i) = 0,1,$ and $2$ represents player $i$ choosing strategy $\emptyset,\{1\},$ and $\{2\}$, respectively. It follows immediately that $A_t(i) \!= \!0$ for all $i\!\in\! \Omega\setminus\Pt$ and $P(\At \!=\!0) \!= 1\!-\!\Pgo$. The proportion of the players that choose each resource in $\Rcal$ is denoted by $\nAt := {[P(\At \!=\! 1)\;P(\At \!= \!2)]^\top\!\!} \in \! \Rnn^2$ and follows $\mathbf{1}^\top \nAt = \Pgo$.

%, i.e., chooses a strategy in $\tilAi := \{\!\{1\},\{2\}\!\}$

The weight of the players is time-invariant and represented by a r.v.\ $W >0$. We may not assume that the range of $W$ is bounded, but we do assume that the overall weight of the population is finite, i.e., $\EV[W] = \int_{\Omega} W \dint P< \infty$. Denote the proportion of the cumulative weight of the players that choose each of the resources in $\Rcal$ by $\wAt \in \Rnn^2$, whereby $\wAt_r \!:= \!\int_{\Omega} P(\At \!=\! r|W)W\dint P / \EV[W]$ for $r\in \Rcal$. We assume that $P_t$ and $W$ are independent, thus it follows that $\mathbf{1}^\top \wAt \!=\! \Pgo$. The players are competing against each other to minimize their own discomfort. We represent the discomfort perceived by a player that chooses resource $r\in \Rcal$ by a player-invariant latency function $l_r\!:\![0,1]\to \Rnn$, which is a strictly increasing continuous function of $\wrAt$. We also write $\mathbf{l}(\wAt) := [l_1(\wAt_1)\; l_2(\wAt_2) ]^\top$.

%\footnote{We consider the conditional probability $P(\At = r|W)$ to be itself a r.v., which is well-defined according to \cite[Chap.~13]{Rosenthal2006}, to circumvent conditioning on an event of probability 0.}

Each instance of the aforementioned game is a weighted singleton congestion game, i.e., each strategy of the players is comprised of a single resource and the latency incurred in using each resource is a function of the cumulative weight of the players that choose that resource only. Notice that without any pricing policy, repeated instances of this game are decoupled. 

\subsection{Artificial Currency Mechanism}

In this section, we propose an AC mechanism for weighted congestion games. Each player is endowed with a wallet of an AC, exclusively for this allocation setting, which cannot be traded or bought with money. The players' level of AC at time $t$ is represented by a r.v.\  $K_t \geq 0$. A generic AC pricing policy maps, at each instant and for each player, the chosen strategy to a payment of the AC. In this paper, we consider a significantly simpler particular case, which (i)~is time-invariant; (ii)~prices the strategies individually for each player, depending on their weight only; and (iii) does not price the empty strategy. Formally, it is denoted by $\pi: \Rp \times \{0,1,2\} \to \R$, which maps a weight $w\in \Rp$ and a choice $a $ to a payment and is given by $\pi(w,a) = \mathbf{p}_{a}(w)$, if $ a \in \Rcal$, and $\pi(w,a) = 0$, if $a = 0$, where $\mathbf{p}:\Rp \to \R^2$ is a  time- and player-invariant piecewise-continuous function that maps a player's weight to a vector of AC payments for choosing each of the resources.\footnote{Piecewise continuity is enforced only to ensure that $\mathbf{p}(W, \At)$ is itself a r.v., as implied by \cite[Propositoin~3.1.8]{Rosenthal2006}.}

Note that the set of available strategies for a participating player is now constrained by their AC level. Specifically, at time $t$ a player $i\in \Omega$ must choose a strategy they can afford, i.e., $\At(i) \in \Ait$, where $\Ait := \{r\in\Rcal: K_t(i) \geq \pi(W(i),r)\}$ for all $i\in \Pt$ and $\Ait = \{0\}$, for all $i\in \Omega\setminus\Pt$ . The AC level of the players is, thus, updated according to $K_{t+1} = K_t - \pi(W,A_t)$. As a consequence of the AC level dynamics, successive instances of the underlying congestion game are now coupled. 

%As a technical remark, we enforced that $\mathbf{p}$ has to be piecewise-continuous so that, if  $K_t$ and $W$ are r.v.\, then $K_{t+1}$ is also an r.v.\ (see \cite[Chap.~3]{Rosenthal2006} for details). 

\subsection{Players' Decision Model}

The coupling between successive instances of the resource allocation shapes the players' decision behavior, which yields a new game that we call the \emph{transactive} game. Indeed, players are also playing against their future selves when deciding to spend AC to choose a resource with a lower latency or earn AC to afford the resource with a lower latency in future instances of the game. 

When choosing a resource at time $t$, a player $i$, under their bounded rationality, ponders: (i)~the latency at time $t$ perceived in the strategy decision $\At(i)$; and (ii)~the latency that follows from future constraints on the strategy space due to the AC constraints. We denote the cost that accounts for both of these components perceived by player $i$ at time $t$  by choosing $a\in \Ait$ given the aggregate decisions of the population $\wAt$ by $c_{\wAt}^{a}(i)$.  As a consequence, the players' best response strategy is given by $\At(i) \in \argmin_{a \in \Ait} c_{\wAt}^{a}(i)$. The first component is modeled through $\{l_r\}_{r\in \Rcal}$. On top of that, since the perception of latency for a player may vary from time to time, we consider that the players have an urgency, borrowing the term from a mobility setting, which weights the latency of choosing a resource at time $t$. The urgency of a player at $t = \{0,1,\ldots\}$ are i.i.d.\ r.v.s $U_t\geq 0$, which are independent to all other r.v.s and whose cumulative distribution function (c.d.f.) is continuous. In this paper, for a participating player $i\in \Pt$, we model $ c_{\wAt}^{r}(i)$ as %is denoted by $F_U$ and
\begin{equation}\label{eq:decision_cost}
	\begin{split}
	 \!\!c_{\wAt}^{r}\!(i) \!=\!\! \min_{\bar{\mathbf{y}}\in \Rnn^{2}} & U_t(i)l_r(\wAt_r) + \EV[U_t]\Pgo T\bar{\mathbf{y}}^\top \mathbf{l}(\wAt)\\
		\mathrm{s.t.} \;\;& \!\!\mathbf{1}^\top \bar{\mathbf{y}} = 1\\
		\phantom{\mathrm{s.t.}} \;& \!\!K_t(i) \!-\! \mathbf{p}_r(W(i)\!) \!-\! \Pgo T\bar{\mathbf{y}}^\top\! \mathbf{p}(W(i)\!) \geq 0,\!\!
	\end{split}
\end{equation}
where $T \in \N$ is a player-invariant decision horizon. Notice that \eqref{eq:decision_cost} accounts for the future perceived latency and future payments in an average fashion. This decision model has been used in previous works, thus the reader is referred to \cite{SalazarPaccagnanEtAl2021,PedrosoHeemelsEtAl2023} for a more thorough analysis. Notice that, given r.v.s $W$, $P_t$, $K_t$, $U_t$, there may be multiple or no decision outcomes $A_t$ that are characterized by \eqref{eq:decision_cost}. In what follows, we characterize outcomes of interest, and in Section~\ref{sec:well-posedness} we establish their existence and essential uniqueness.

\subsection{Efficiency}

At each time $t$, we model the decision outcome of the population as a Nash Equilibrium (NE) of the transactive game, which is defined in what follows. Intuitively, at  time $t$, a decision outcome $\At$ is a NE if no player can individually change their strategy to obtain a better outcome. Henceforth, we are only concerned with decision outcomes that are NE.

\begin{definition}[NE]
	A decision outcome $A_t : \Omega \to \{0,1,2\}$ is a NE, if $\forall i\in \Omega\; \forall a \in \Ait\;\; c_{\wAt}^{\At(i)} \leq c_{\wAt}^{a}$. %under policy $\mathbf{p}$ 
\end{definition}

From the macroscopic system-level perspective, there is a cost associated with the aggregate use of the resources (e.g., average perceived latency by the players), which we represent as a function $C: \R^2 \to \Rnn$ of $\wAt$. Due to the self-interested behavior of the players, the NE may be far from the system optimum (SO). To characterize the inefficiency of equilibria, we employ the price of anarchy (PoA) \cite{KoutsoupiasPapadimitriou1999}, which is defined at time $t$ by
\begin{equation*}% \vspace{-0.1cm}
	\PoA_t := {\max_{\At\in \AtNE} C(\wAt)}\big/{\min_{\At\in \AtNE} C(\wAt)},
\end{equation*}
where $\AtNE$ denotes the set of NE of the transactive game at time $t$. Notice the PoA attains a minimum of 1. 

\subsection{Fairness: Equality and Equity}

The number of times that a player participated in the game until time $t$ is a r.v.\ $N_t := \sum_{\tau = 0}^tP_\tau$. The average perceived latency until time $t$ is a r.v.\ $L_t$ given by
\begin{equation*}
	L_t(i) := \left(\sum\nolimits_{\substack{\tau = 0\\P_\tau(i) = 1}}^{t} l_{A_\tau}(\mathbf{w}^{A_\tau})\right)/N_t(i),
\end{equation*} 
if $N_t(i) >0$, and $L_t(i) = 0$, otherwise, for all $i\in \Omega$. In this paper, we define inequity and inequality as the standard deviation of $L_t$ and $L_t/W$, respectively. Formally, $\InEqt_t^2 := \Var[L_t]$ and  $\InEql_t^2 := \Var[L_t/W]$.

\subsection{Artificial Currency Pricing Policy Design Problem}

We are now ready to define the artificial currency pricing policy design problem. We aim at designing $\mathbf{p}$, which is a function of the players' weights, such that inequity or inequality are minimized, while still converging to the SO.

\begin{problem}
	Design $\mathbf{p}$ such that the limit $\lim_{ t\to \infty}  \InEqt_t$ (or $\lim_{t \to \infty} \InEql_t$) exists and is minimized, subject to the condition that $\lim_{t \to \infty} \PoA_t$ exists and is unitary.
\end{problem}

% !TeX spellcheck = en_US
\section{Pricing Policy Design}\label{sec:design}

Before beginning with the pricing policy design approach, we select a minimizer $\wstar$ of the societal cost, i.e., $\wstar \in \argmin_{\mathbf{w} \in \Rnn^2} C(\mathbf{w}) \; \mathrm{s.t.} \; \mathbf{1^\top \mathbf{w} = \Pgo}$. We also write $\mathbf{l}^\star := \mathbf{l}(\mathbf{w}^\star)$. Henceforth, we make four assumptions: (i)~$\wstar_1 > 0$ and $\wstar_2 > 0$, otherwise at the SO all players use one and the same resource; (ii)~$l_1(\wstar_1) \neq l_2(\wstar_2)$ and we label the two resources such that $l_1(\wstar_1) < l_2(\wstar_2)$; (iii)~$\mathbf{p}_2(w) < \mathbf{p}_1(w) \;\forall w\in \Rp$, which implies that $\mathbf{p}_2(w) \leq 0\; \forall w\in \Rp$ since a participating player must have at least one strategy available; and (iv)~$C([w\; \Pgo-w]^\top)$ is Lipschitz continuous w.r.t.\ $w \in [0,\Pgo]$ with Lipschitz constant $L_C >0$ w.r.t.\ the absolute difference metric.

We also introduce the rational approximation function $\rat_\delta$, which is instrumental in the design procedure. For fixed $\delta >0$, the function $\rat_\delta : \Rnn \to \Q_{\geq0}$ is a piecewise-constant function, which is constant on the intervals $\Acal_j := ((j-1)\delta,j\delta]$ such that $\rat_\delta(x) := \sum_{j=1}^\infty q_j\mathbf{1}_{\Acal_j}(x)$, where $q_j \in \Qp$ satisfies $|q_j-x|<\delta\; \forall x\in \Acal_j$.

\subsection{Well-posedness}\label{sec:well-posedness}

In this section, we establish the existence and essential uniqueness of the NE of the decision outcome of the transactive game, which are necessary to establish the well-posedness of the design problem. Specifically, we prove that a NE exists for any AC level and urgency of the population and that the cumulative weight of the players choosing each of the resources at the NE is unique. 

%Second, we detail important definitions and notation concerning the AC dynamics.

%[Existence and Essential Uniqueness of NE]
\begin{theorem}\label{thm:NE}
	At time $t$, given $W$, $K_t$, $U_t$, and a pricing policy $\mathbf{p}: \Rp \to \R^2$, the set of NE is nonempty and essentially unique, i.e., $\AtNE \neq \emptyset$ and $\forall \At^1, \At^2 \in \AtNE \; \mathbf{w}^{\At^1} = \mathbf{w}^{\At^2}$. Furthermore, if $l_1(\mathbf{w}^{\At^1}) < l_2(\mathbf{w}^{\At^1})$, then $\At^1$ and $\At^2$  are unique up to a set of probability 0.
\end{theorem}
\begin{proof}
	\ifextendedversion
	See Section~\ref{sec:proof_NE} in Appendix.
	\else
	See the extended version of this paper~\cite{PedrosoHeemelsEtAl2024Extended}.
	\fi
\end{proof}

Henceforth, we will be exclusively concerned with sequences of r.v.s for which the decision outcome NE $A_t$ abides by $l_1(\mathbf{w}^{\At}) < l_2(\mathbf{w}^{\At})$ for all $t\in \Nz$, which is where the SO lies. By Theorem~\ref{thm:NE}, in this region, at time $t$ the NE decision outcome $\At$ is completely defined (up to a set of probability 0) by $W$, $K_t$, $U_t$, and $\mathbf{p}$. In the sequel, we establish that there are initial conditions for which $\colK$ remains in this region and converges in distribution to a unique AC distribution.

%\begin{definition}[Asymptotically Stable AC Equilibrium]
%	
%\end{definition}

\subsection{Design for Equity}

In this section, we propose a pricing policy that achieves maximum equity as $t\to \infty$, while still ensuring convergence to the SO. Given that maximum equity is achieved whenever all players endure the same average latency irrespective of their weight, then it is natural to consider a pricing policy that does not depend on the players' weights. If the initial AC level distribution $K_0$ and $W$ are independent, then $A_t$ and $W$ are independent, and $\wAt = \nAt$. Generalizing the approach in \cite{SalazarPaccagnanEtAl2021}, one could speculate that prices for which the expected AC level of the population remains constant at the SO, i.e., $\mathbf{p}^\top\wstar = 0$, achieve these goals. The following result shows that, indeed as $t\to \infty$ a policy inspired  that reasoning achieves perfect equity and the NE converges arbitrarily close to the SO.

\begin{theorem}\label{thm:Eqt}
		For a sufficiently small $\epsilon >0$, consider a pricing policy
	\begin{equation*}
		\mathbf{p}(w) = {S[\rat_\delta(\wstar_2/\wstar_1)\, -\!1]^\top},
	\end{equation*}
	where $\delta = (\epsilon\Pgo C(\wstar)/(\wstar_1(L_C\wstar_1 +\epsilon C(\wstar))$ and $S \in \Q_{>0}$. Then, there exists an initial AC level distribution for which $\PoA_t$ and $\InEqt_t$ converge and $\lim_{t \to \infty} \PoA_t \leq 1+ \epsilon$ and $\lim_{t \to \infty} \InEqt_t = 0$. %Furthermore, if $\EV[1/W^2]\leq \infty$, then $\InEql_t$ converges and $|\lim_{t \to \infty} \InEql_t - ({\mathbf{l}^\star}^\top \mathbf{w}^\star/\Pgo)(\EV[1/W^2]-\EV[1/W]^2)| \leq \epsilon(\mathbf{l}^\star_2 -\mathbf{l}^\star_1)/(L_c \Pgo)$.
\end{theorem}
\begin{proof}
	\ifextendedversion
	See Section~\ref{sec:proof_Eqt} in Appendix.
	\else
	See the extended version of this paper~\cite{PedrosoHeemelsEtAl2024Extended}.
	\fi
\end{proof}

\begin{remark}
	From Theorem~\ref{thm:Eqt}, convergence is only guaranteed for particular initial AC level distributions, which are required, in the proof, to be sufficiently close to the limit AC distribution. Nevertheless, in practice, convergence to the SO is seen for all initial conditions tested by the authors.
\end{remark}

\begin{remark}
	Notice that the optimal pricing policy in Theorem~\ref{thm:Eqt} is invariant to a positive scaling $S \in \Q_{>0}$. This is in order, since the value of the AC is arbitrary, i.e., it is not pegged to any monetary or value-of-time point of reference. Indeed, the ratio between the prices alone defines the aggregate equilibria.
\end{remark}

\subsection{Design for Equality}

In this section, we propose a pricing policy that achieves the best possible equality as $t\to \infty$. We make one further assumption, that the support of $W$, i.e., the smallest closed interval $\Wcal = [\wmin, \wmax] \subseteq \R$ such that $P(W\in \Wcal) = 1$, is bounded and $\wmin >0$. For simplicity of the notation, we also admit that $\wmin \leq W \leq \wmax$, which could be lifted trivially. The intuition behind the proposed approach is to partition the set of players in infinitesimal weight brackets. Then, constant prices are designed for each bracket such that a prescribed proportion of the players in the bracket chooses strategy $\{1\}$. The proportions that are prescribed in each bracket are designed such that (i)~the weighted average perceived latency in all brackets is the same (or as close as possible) at the SO; and (ii)~the overall prescribed cumulative weight among all brackets corresponds to the SO.

Specifically, we desire to first design a r.v.\ $n_1(W,\theta)$, where 
$n_1: \Wcal \times \Rp \to [0, \Pgo]$ maps a weight $w\in \Wcal$ and a scalar parameter $\theta \in \Rp$ to a prescribed proportion of choosing strategy $\{1\}$. Parameter $\theta$ is used exclusively to parameterize the family of r.v.s $n_1(W,\cdot)$ of interest, which is defined below. We also define $n_2: \Wcal \times \Rp \to [0; \Pgo]$ as $n_2 = \Pgo-n_1$. Then, prices can be chosen so that ${P(\At \!=\! 1 |W)}$ converges to the prescribed r.v.\ $n_1(W,\theta^\star)$, where $\theta^\star$ is designed to achieve maximum equality. Null inequality at the SO, would mean that for some constant $C_\theta \!\in\! \Rp$, which we parameterize with $\theta$, $n_1(w,\theta)$ satisfies
\begin{equation}\label{eq:cond_perf_eql}
	C_\theta = \frac{n_1(w,\theta)\lstar_1 + (\Pgo-n_1(w,\theta))\lstar_2}{w\Pgo}
\end{equation}
for any $w\in \Wcal$. Henceforth, we parameterize $C_\theta$ according to $C_\theta = (\wstar_1\lstar_1 +(\Pgo-\wstar_1)\lstar_2)/(\theta \Pgo)$, which uniquely defines any $C_\theta \in \Rp$ with a $\theta \in \Rp$. Notice that the individual players' average perceived latency at the SO is upper bounded by $\lstar_2$ and lower bounded by $\lstar_1$, which corresponds to always choosing $\{2\}$ and $\{1\}$, respectively. As a result, if $\Wcal$ is wide enough, i.e., if there exist $w_1,w_2\in \Wcal$ such that $\lstar_1w_2 > \lstar_2w_1$, then it is not possible to satisfy \eqref{eq:cond_perf_eql} for every $w\in \Wcal$ such that $0\leq n_1(w,\theta)\leq \Pgo$. Therefore, for some $C_\theta$, maximum equality is achieved when $n_1(w,\theta)$ is defined by \eqref{eq:cond_perf_eql}, if feasible, and to the closest limit otherwise, i.e.,
\begin{equation*}
n_1(w,\theta) = \begin{cases}
	\Pgo, & \frac{w}{\theta}\!-\!1 \leq  -\frac{\Pgo-\wstar_1}{\xi-\wstar_1}\\
	\!-\frac{w}{\theta}(\xi\!-\!\wstar_1) +\xi, & -\frac{\Pgo-\wstar_1}{\xi-\wstar_1} \!< \!\frac{w}{\theta}\!-\!1\! <\! \frac{\wstar_1}{\xi-\wstar_1}\\
	0, &  \frac{w}{\theta}\!-\!1 \geq  \frac{\wstar_1}{\xi-\wstar_1},
\end{cases}	
\end{equation*} 
where $\xi:= \Pgo \lstar_2/(\lstar_2-\lstar_1)$.

Now, one must find the optimal $\theta^\star$ such that the overall cumulative prescribed weight corresponds to the SO, i.e., 
\begin{equation}\label{eq:design_theta}
	\wstar_1 = \int_\Omega n_1(W,\theta^\star)W \dint P/ \EV[W].
\end{equation} 
In the following result, we establish that there exists at least one solution $\theta^\star \in \Rp$ to \eqref{eq:design_theta} and that, if it is not unique, all solutions prescribe the same optimal equality.

\begin{lemma}\label{lem:sol_theta}
	If the support of $W$ is $\Wcal = [\wmin, \wmax] \subset \Rp$, then  \eqref{eq:design_theta} has at least one solution $\theta^\star \in \Rp$. Furthermore, any solution $\theta^\star \in \Rp$ prescribes the same inequality.
\end{lemma}
\begin{proof}
		\ifextendedversion
		See Section~\ref{sec:proof_sol_theta} in Appendix.
		\else
		See the extended version of this paper~\cite{PedrosoHeemelsEtAl2024Extended}.
		\fi
\end{proof}

\begin{remark}
	The condition for the optimality of the parameter $\theta$ in \eqref{eq:design_theta} and the existence of a solution in Lemma~\ref{lem:sol_theta} is stated for a general weight distribution, as long as the support of $W$ is bounded. Nevertheless, it is interesting to remark two particular cases: (i)~if $W$ is discrete and takes a finite number of values $\{w_i\}$ each with probability $p_i$, then \eqref{eq:design_theta} becomes
	\begin{equation*}
		\wstar_1 \sum\nolimits_i w_ip_i = \sum\nolimits_i n_1(w_i,\theta^\star)w_ip_i;
	\end{equation*}
	and (ii)~if $W$ is continuous with Riemann-integrable probability density function (p.d.f.) $f_W$, then \eqref{eq:design_theta} becomes
	\begin{equation*}
		\wstar_1 \int\nolimits_{\Wcal} wf_W(w) \dint w=  \int\nolimits_{\Wcal} n_1(w,\theta^\star)wf_W(w) \dint  w.
	\end{equation*}
	Finding $\theta^\star$ amounts, for both of these cases, to numerically computing a root of a real-valued function in one variable.
\end{remark}

%Now that the r.v.\ $n_1$,  which represents the prescribed proportion of players that choose strategy $\{1\}$ given their wight, was been optimally design.

Now that the r.v.\ $n_1$ has been optimally designed, the final step is to design the pricing policy $\mathbf{p}$ that guarantees that $n_1$ is prescribed as $t\to \infty$. The following result shows that indeed there exists a policy such that the PoA and the equality converge arbitrarily close to the optimal values.

\begin{theorem}\label{thm:Eql}
	Consider a solution $\theta^\star$ to \eqref{eq:design_theta}. For a sufficiently small $\epsilon >0$, consider a pricing policy
	\begin{equation*}
		\mathbf{p}(w) = 
		\begin{cases}
				S\left[\rat_\delta\!\left(\frac{n_2(w,\theta^\star)}{n_1(w,\theta^\star)}\right)\; -\!1\right]^\top\!\!, & \frac{w}{\theta^\star} \leq 1\\
				S\rat_\delta\!\left(\frac{\wstar_2}{\wstar_1}\right) \left[1 \;-\!\rat_\delta\!\left(\frac{n_1(w,\theta^\star)}{n_2(w,\theta^\star)}\right)\right]^\top\!\!, & \frac{w}{\theta^\star} > 1,
		\end{cases}
	\end{equation*}
	where $\delta = \epsilon\Pgo C(\wstar)/L_C$ and $S \in \Q_{>0}$. Then, there exists an initial AC level distribution for which $\PoA_t$ and $\InEql_t$ converge and $\lim_{t \to \infty} \PoA_t \leq 1+ \epsilon$ and $|\lim_{t \to \infty} \InEql_t -{\InEql^\star}| < \sqrt{D\epsilon}$, where $\InEql^\star$ is the optimal inequality and $D\in \Rp$ is a constant (which is defined in the proof).
\end{theorem}
\begin{proof}
	\ifextendedversion
	See Section~\ref{sec:proof_Eql} in Appendix.
	\else
	See the extended version of this paper~\cite{PedrosoHeemelsEtAl2024Extended}.
	\fi
\end{proof}

\begin{remark}
Notice that a digital representation of a real number is no more than a rational approximation function similar to $\rat_\delta$. As a result, a digital implementation (with enough precision) of the pricing policy in Theorems ~\ref{thm:Eqt} and~\ref{thm:Eql} may omit $\rat_\delta$ altogether. 
\end{remark}

Notice that the prices designed to maximize equity are identical to the prices designed for unweighted congestion games (see~\cite{SalazarPaccagnanEtAl2021}) and do not depend on the weights. However, the prices designed to maximize equity do.

% !TeX spellcheck = en_US
\section{Numerical Results}\label{sec:results}
 In this section, numerical results are presented as an illustrative example for a finite population of $10^3$ users. We consider that $\Pgo = 0.95$, the urgency r.v.s have a uniform distribution on the interval $[0,2]$ and the users' prediction horizon is $T = 4$. The weight r.v. has a truncated normal distribution on the interval $\Wcal = [1/2, 3/2]$ with unitary expected value and a standard deviation of $0.15$ weight units. The latency functions have the form $l_j(w) = \mathbf{l}_j^{\mathbf{0}}(1+\alpha (w/\boldsymbol{\kappa}_j)^\beta)$ where $\alpha = 0.15$, $\beta = 4$, $\mathbf{l^0} = [1 \;2]^\top\!$, and $\boldsymbol{\kappa} = [1/2 \;2/3]^\top\!$. The societal cost is the average latency, i.e., $C(\mathbf{w}) = \mathbf{w}^\top \mathbf{l}(\mathbf{w})$. Rounded to four significant digits, one obtains $\wstar = [0.5596 \;0.3904]^\top$ and $\mathbf{l}(\wstar) = [1.235 \; 2.035 ]^\top$. We chose $S = 10$ and the AC level of the population was initialized randomly from a uniform distribution on the interval $[5S,10S]$, which is independent from $W$.
 
\begin{figure}[ht!]
	\begin{subfigure}{\linewidth}
		\centering
		\includegraphics[width = .98\linewidth,trim={0 .0cm 0 0},clip=true]{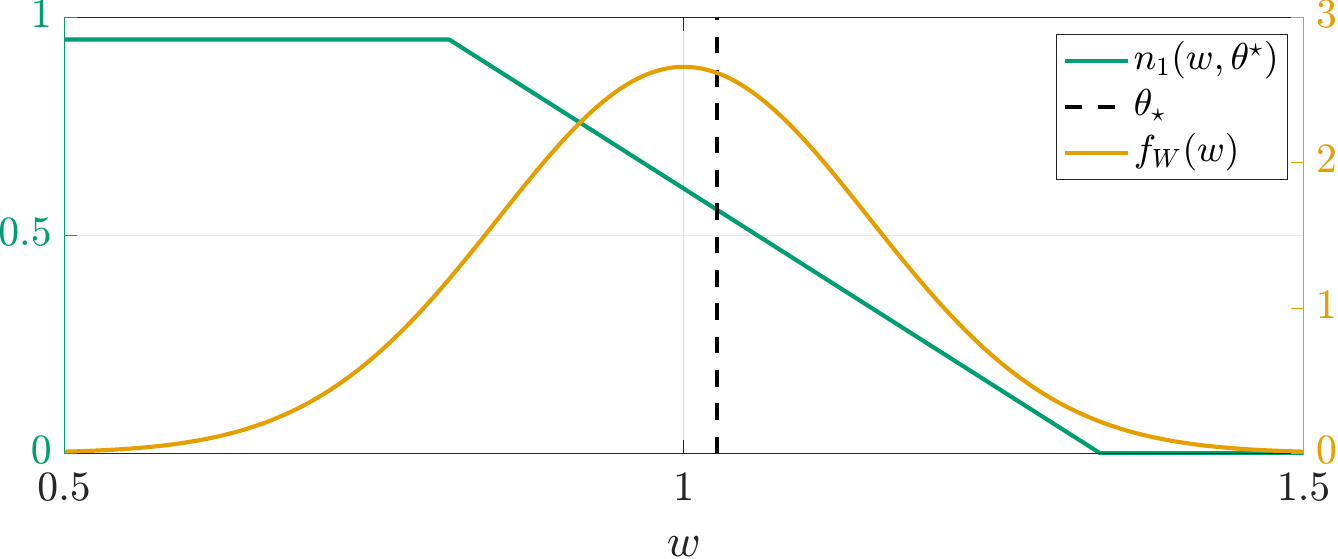}
		\vspace{-0.1cm}
		\caption{$n_1(w,\theta^\star)$ and p.d.f.\ of $W$ as function of the weight.}
		\label{fig:n1}
	\end{subfigure}\vspace{0.5cm}  \\ 
	\begin{subfigure}{\linewidth}
		\centering
		\includegraphics[width = .98\linewidth,trim={0 .0cm 0 0},clip=true]{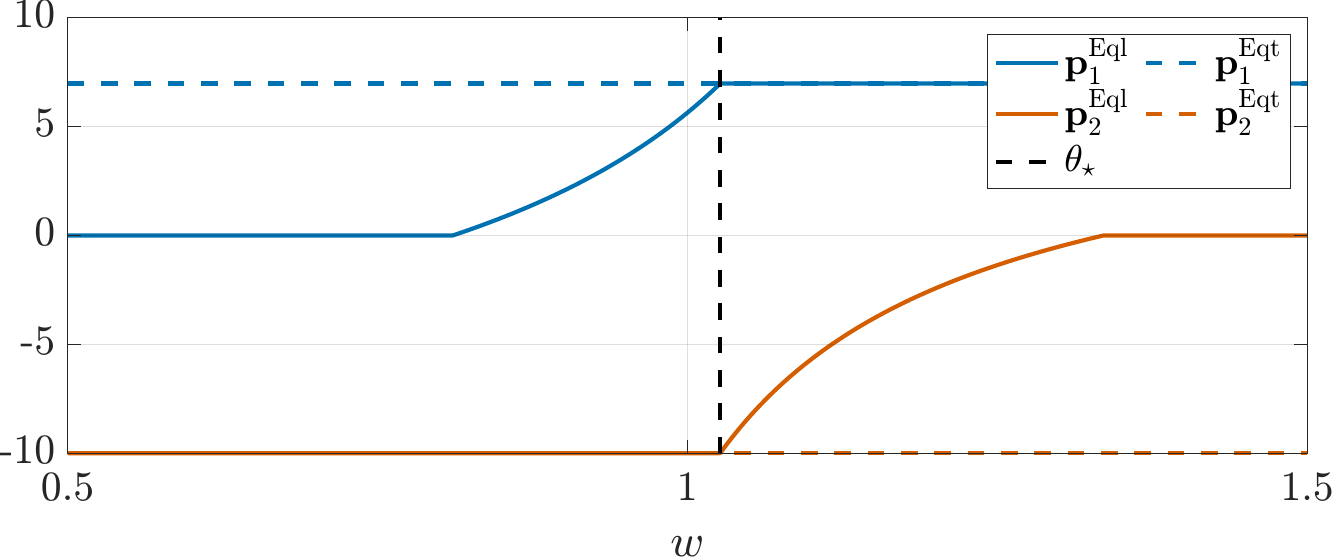}
		\vspace{-0.1cm}
		\caption{Optimal prices as a function of the weight.}
		\label{fig:prices}
	\end{subfigure}\vspace{0.5cm}  \\ 
	\begin{subfigure}{\linewidth}
		\centering
		\includegraphics[width = .98\linewidth,trim={0 0cm 0 0},clip=true]{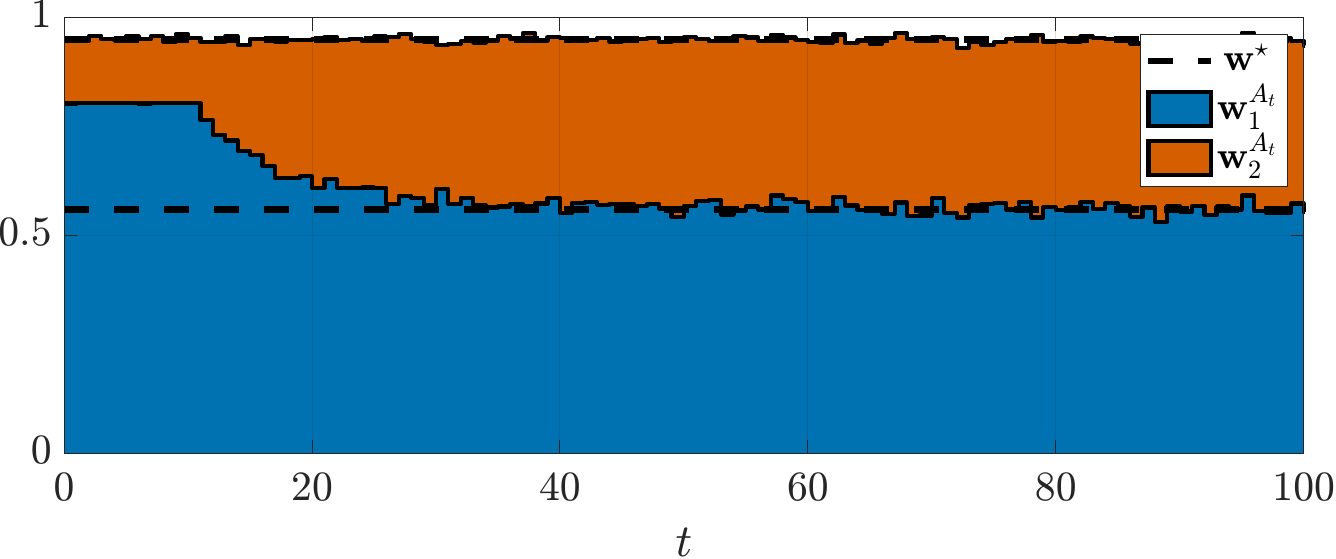}
		\vspace{-0.1cm}
		\caption{Aggregate decision for prices designed for equality.}
		\label{fig:decision}
	\end{subfigure}\vspace{0.5cm} \\
	\begin{subfigure}{\linewidth}
		\centering
		\includegraphics[width = .98\linewidth,trim={0 0cm 0 0},clip=true]{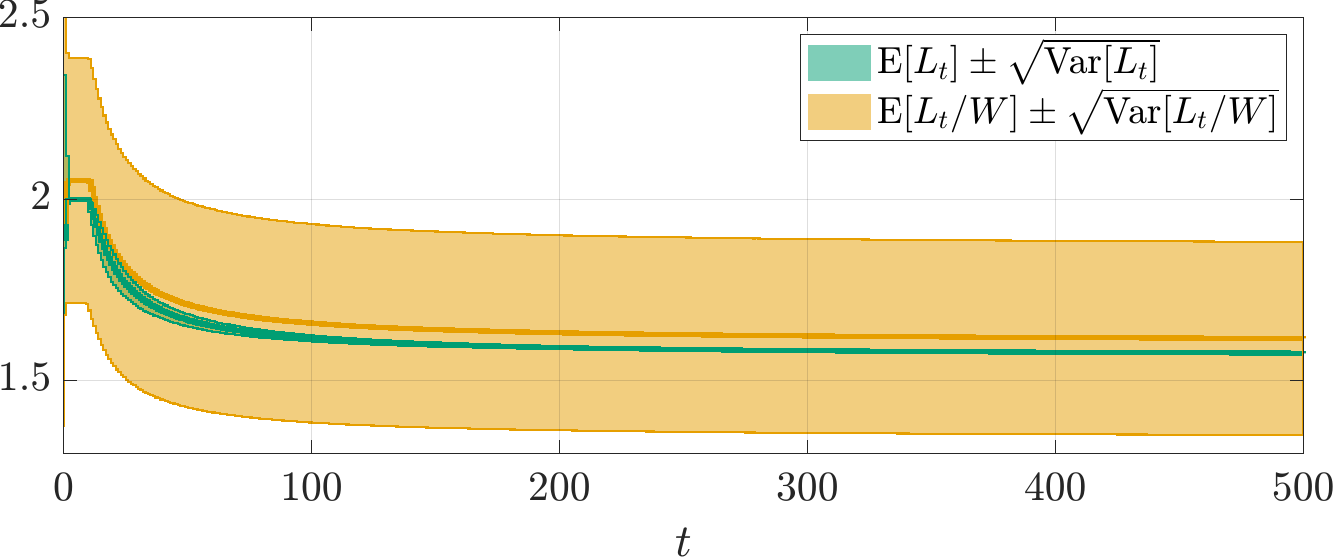}
		\vspace{-0.1cm}
		\caption{Average perceived latency until $t$ (designed for equity).} %Evolution of the relative difference between societal cost and the SO.
		\label{fig:L_Eqt}
	\end{subfigure}\vspace{0.5cm} \\
	\begin{subfigure}{\linewidth}
		\centering
		\includegraphics[width = .98\linewidth,trim={0 0cm 0 0},clip=true]{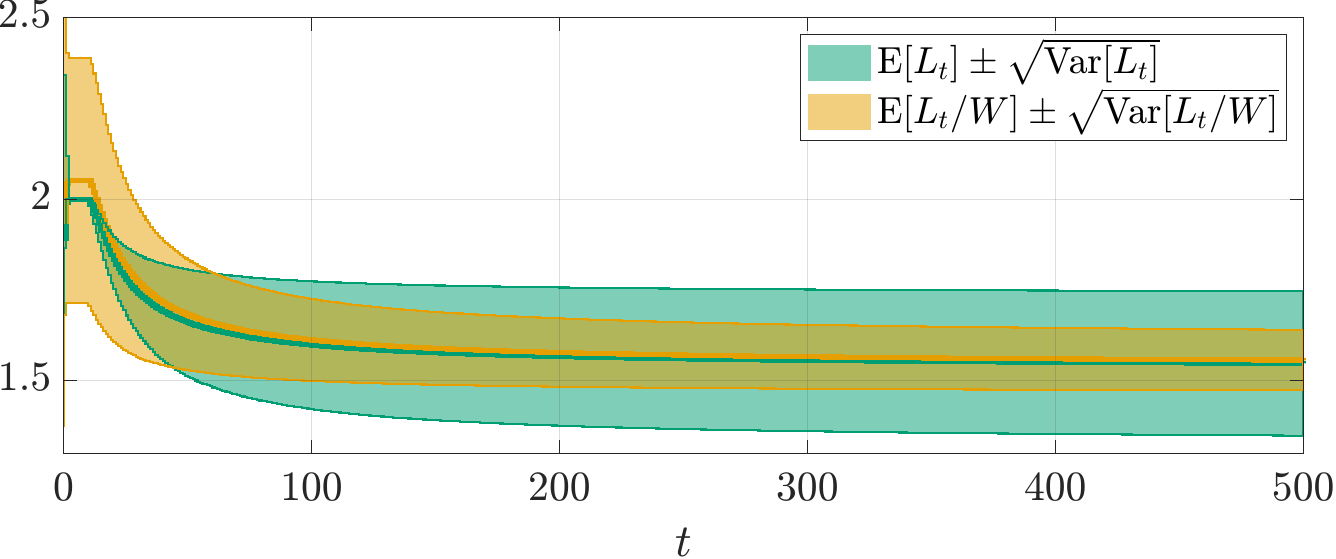}
		\vspace{-0.1cm}
		\caption{Average perceived latency until $t$ (designed for equality).} 
		\label{fig:L_Eql}
	\end{subfigure}	\vspace{-0.6cm} \\
	\caption{Numerical simulation results with comparison between the AC policy design for equity and equality.}
	\label{fig:num_sim}
\end{figure}

 In Fig.~\ref{fig:n1}, the p.d.f.\ of $W$ and $n_1(w,\theta)$, for $\theta^\star = 1.027$, are depicted as a function of the weight. The optimal prices designed to optimize for equity and equality are both shown in Fig.~\ref{fig:prices} as a function of the weight. In Fig.~\ref{fig:decision}, the evolution of the aggregate user decisions with prices designed to optimize equality is depicted. Even though the initialization of the AC level was arbitrary, we see that the aggregate decision  converges to the SO. The evolution with prices designed for equity is quite identical and also converges to the SO. For the prices designed for equity, Fig.~\ref{fig:L_Eqt} depicts the evolution of the average perceived unweighted and weighted latency until time $t$ and the respective bounds on the standard deviation (across users), which correspond to $\InEqt_t$ and $\InEql_t$, respectively. Fig.~\ref{fig:L_Eql} depicts the same for prices designed for equality. As expected, from Fig.~\ref{fig:L_Eqt}, the prices designed for equity achieve perfect equity, but result in a large inequality. From Fig.~\ref{fig:L_Eql}, the prices designed for equality do reduce the inequality substantially compared to Fig.~\ref{fig:L_Eqt}, but cannot achieve perfect equality because the support of $W$ is wide in this example.
 
 Due to space limitations, some details regarding the numerical pricing design and the simulation were omitted. However, a MATLAB implementation as well as additional simulation results are openly available in an open source repository at \href{https://fish-tue.github.io/AC-weighted-eqt-eql}{\tt \small https://fish-tue.github. io/AC-weighted-eqt-eql}.

% !TeX spellcheck = en_US
\section{Conclusions}\label{sec:conclusions}

In this paper, we explore the equity vs. equality dichotomy for the first time in the literature of mechanism design. We design two optimal artificial currency incentive schemes that maximize equity and equality in repeated weighted congestion games, while still achieving system-optimal performance. We consider a policy that always allows a user to afford at least one resource to satisfy their needs. In that case, it can be concluded that there exists a policy that achieves perfect equity. However, if the weight distribution has a wide support, it may not be possible to design a policy to achieve perfect equality. It is noticeable that the design procedure as well as the aggregate outcome are dramatically distinct for each of the fairness criteria. Both optimal policies achieve convergence to the system optimum. Future research endeavors will include leveraging the potential of this framework to a mobility-on-demand management problem, whereby users choose between ride-pooling or traveling alone. In this case, trips with different distances contribute differently to congestion. The latency functions can be given by a macroscopic ride-pooling model~\cite{PaparellaPedrosoEtAl2024b}.
% !TeX spellcheck = en_US

%\section*{Code Availability Statement}
%A MATLAB implementation of the methods and simulations presented in this paper are openly available in an open-source repository available at {\small\texttt{\url{https://fish-tue.github.io/ }}}.

\section*{Acknowledgment}
We thank Dr.\ I.\ New for proofreading the paper.

%\appendices
\ifextendedversion
% !TeX spellcheck = en_US
\section*{Appendix: Proofs of Section~\ref{sec:design}}

The proofs in this appendix require introducing more notation. The floor (ceiling) of a real number $x$, which is the greatest integer less than or equal to $x$ (smallest integer greater than or equal to $x$), is denoted by $\lfloor x\rfloor$ ($\lceil x \rceil$). The greatest common divisor between two natural numbers $a,b \in \N$ is denoted by $\gcd(a,b)$. 

\subsection{Proof of Theorem~\ref{thm:NE}}\label{sec:proof_NE}

Before proceeding, is it useful to introduce a lemma. 

\begin{lemma}\label{lem:brs}
	Consider a player $i\in \Omega$ and define $k_{(1,2)}:= \mathbf{p}_1(W(i))$, $k_{(2,1)}:=\max(\mathbf{p}_1(W(i)),\mathbf{p}_2(W(i))+\Pgo T\mathbf{p}_1(W(i)))$, $k_{(1,1)}:= (\Pgo T\!+\!1)\mathbf{p}_1(W(i))$, and 
	\begin{equation*}\small
		\gamma(k) \!:= \!\begin{cases}
			+\infty, &\!\!\! k \!< \! k_{(1,2)}\\
			1, & \!\!\! k_{(1,2)} \!\leq \! k \!\leq \! k_{(2,1)}\\
			\!\!(k_{(1,1)}\!-\! k)/(\mathbf{p}_1(W(i)\!) \!-\! \mathbf{p}_2(W(i)\!)\!), &\!\!\! k_{(2,1)} \!\!<\!k \!< \!k_{(1,1)}\\
			0, & \!\! k \!\geq \!k_{(1,1)}.
		\end{cases}
	\end{equation*}
	The best response strategy (BRS) of $i \in \Omega$, from \eqref{eq:decision_cost}, is
	\begin{equation*}
		\At(i) \! \begin{cases}
		\!\!\begin{cases}
			=2, &  \!\!U_t(i)/\EV[U_t] \!<\! \gamma(K_t(i))\\
			=1, & \!\! U_t(i)/\EV[U_t] \!>\! \gamma(K_t(i))
		\end{cases}\!\!, & \!\!l_1(\wAt_1) \!<\! l_2(\wAt_2)\\[1em]
		\!\!\begin{cases}
				= 2, &  K_t(i) \!<\! \mathbf{p}_1(W(i))\\
				\in \{1,2\}, & K_t(i) \geq  \mathbf{p}_1(W(i))
			\end{cases} \!\!,&\!\! l_1(\wAt_1) \!=\! l_2(\wAt_2)\\
			2 , & \!\!l_1(\wAt_1) \!> \!l_2(\wAt_2)
			\end{cases}
	\end{equation*}
if $i\in \Pt$ and $A_t(i) = 0$, otherwise. If the urgency conditions above are satisfied with equality any $r\in \{1,2\}$ is a BRS.
%		\begin{equation*}
%		\At(i) = \begin{cases}
%			2, &  u_t(i)/\EV[U_t] < \gamma(K_t(i))\\
%			1, & u_t(i)/\EV[U_t] > \gamma(K_t(i))\\
%		\end{cases}
%	\end{equation*}
%	
%	\begin{equation*}\small
%		r \!= \!\begin{cases}
%			2, &\!\!\! k_t(i) < \mathbf{p}_1(w(i))\\
%			\!\!\begin{cases}
%				2, & \!\!\! u_t(i) \!\! < \!\EV[U]\\
%				1, & \!\!\! u_t(i) \!>\!  \EV[U]
%			\end{cases},& \!\!\!  \mathbf{p}_1(w(i)\!) \leq k_t(i) \!\leq\! k_{(2,1)}\\
%			\!\!\begin{cases}
%				2, & \!\!\! u_t(i)\! \! < \! \EV[U]\frac{k_{(1,1)}-k_t(i)}{\mathbf{p}_1(\!w(i)\!) - \mathbf{p}_2(\!w(i)\!)}\\
%				1, & \!\!\! u_t(i) \!>\!  \EV[U]\frac{k_{(1,1)}-k_t(i)}{\mathbf{p}_1(\!w(i)\!) - \mathbf{p}_2(\!w(i)\!)}
%			\end{cases}\!\!,& \!\!\! k_{(2,1)} \!\! < \!k_t(i) \!\leq \!k_{(1,1)}\\
%		1, &\!\!\! k_t(i) > k_{(1,1)},
%		\end{cases}
%	\end{equation*}
%	if $l_1(\mathbf{n}^{a_t}_1) < l_2(\mathbf{n}^{a_t}_1)$, 
%	\begin{equation*}
%			r \begin{cases}
%			= 2, &  k_t(i) < \mathbf{p}_1(w(i))\\
%			\in \{1,2\}, & k_t(i) \geq  \mathbf{p}_1(w(i)),
%		\end{cases}
%	\end{equation*}
%	if $l_1(\mathbf{n}^{a_t}_1) = l_2(\mathbf{n}^{a_t}_2)$, and $r =2$, if $l_1(\mathbf{n}^{a_t}_1) > l_2(\mathbf{n}^{a_t}_1)$. If the urgency conditions above are satisfied with equality any $r\in \{1,2\}$ is a best response strategy.\notef{The presentation of this lemma is very ugly and takeas a lot of space. Define a function gamma here like cdc.}
\end{lemma}
\begin{proof}
	The proof of this lemma is a straightforward adaptation of \cite[Theorem~4.1]{SalazarPaccagnanEtAl2021}.
\end{proof}

First, any decision outcome $\At$ such that $l_1(\wAt_1) >  l_2(\wAt_2)$ cannot be a NE. Indeed, since it must be the case that $\wAt_2 < \wstar_2$ and it follows from Lemma~\ref{lem:brs} that the BRS of every player is to choose strategy $\{2\}$, then at least one player can change their decision to obtain a better outcome. Second, notice that, according to Lemma~\ref{lem:brs}, the BRS of every player given a decision outcome $\At$ of the remainder of the population such that $l_1(\wAt_1) <  l_2(\wAt_2)$ does not depend on the magnitude of  $l_1(\wAt_1)$  or  $l_2(\wAt_2)$. Given $W$, $K_t$, $U_t$, and $A_t$ denote the decision outcome that follows from the BRS of every $i \in \Omega$ with $l_1(\wAt_1) <  l_2(\wAt_2)$ by $A^\BRS_t$. Since $F_U$ is continuous, the measure of the set of players whereby the urgency conditions in Lemma~\ref{lem:brs} are satisfied with equality is null. Therefore, $\wAtBRS$ is unique. Then, if $l_1(\wAtBRS_1) <  l_2(\wAtBRS_2)$, it follows that $\AtBRS$ is a NE, because given $\AtBRS$ no player can individually change their decision to obtain a better outcome. Now, to prove the essential uniqueness of any NE $\At$ that abides by $l_1(\wAtBRS_1) <  l_2(\wAtBRS_2)$, take two such NE $\At^1, \At^2$. Since they are NE, $\At^1$ and $\At^1$ must be a BRS of every player given the decision of the remainder of the population in $\At^1$ and $\At^2$, respectively. From the BRS aggregate uniqueness arguments above, since $\At^1$ and $\At^2$ are both the outcome of the BRS of every player, then $\At^1$ and $\At^1$ are unique up to a set of probability 0 and, thus, $\mathbf{w}^{\At^1} = \mathbf{w}^{\At^2}$. Third, if  $l_1(\wAtBRS_1) \geq  l_2(\wAtBRS_2)$, by the continuity and monotonicity of $\{l_r\}_{r\in \Rcal}$, there exists an unique $\wbar$ that abides by $l_1(\wbar_1) = l_2(\wbar_2)$ and $\mathbf{1}^\top\wbar = \Pgo$. As a result, it follows that $\wAtBRS_2 \leq \wbar_2$, which implies that the proportion of the cumulative weight of players forced to choose $\{2\}$ because of the artificial currency constraints is lower that $\wbar_2$, i.e., $P(P_t = 1 \cap K_t \leq \mathbf{p}_1(W)) \leq \wbar_2 $. Therefore, any decision outcome $\At$ whereby all players that are forced to choose $\{2\}$ are allocated to $\{2\}$ and the remaining participating are split between $\{1\}$ and $\{2\}$ such that  $l_1(\wAtBRS_1) =  l_2(\wAtBRS_2)$ is a NE. Essential uniqueness of these NE follows immediately. $\hfill\square$

%\begin{theorem}
%	For a sufficiently small $\epsilon >0$, consider a pricing policy
%	\begin{equation*}
%		\mathbf{p}(w) = {S[\rat_\delta(\wstar_2/\wstar_1)\, -\!1]^\top},
%	\end{equation*}
%	where $\delta = (\epsilon\Pgo C(\wstar)/(\wstar_1(L_C\wstar_1 +\epsilon C(\wstar))$ and $S \in \Q_{>0}$. Then, there exists an initial AC level distribution for which $\PoA_t$ and $\InEqt_t$ converge and $\lim_{t \to \infty} \PoA_t \leq 1+ \epsilon$ and $\lim_{t \to \infty} \InEqt_t = 0$. %Furthermore, if $\EV[1/W^2]\leq \infty$, then $\InEql_t$ converges and $|\lim_{t \to \infty} \InEql_t - ({\mathbf{l}^\star}^\top \mathbf{w}^\star/\Pgo)(\EV[1/W^2]-\EV[1/W]^2)| \leq \epsilon(\mathbf{l}^\star_2 -\mathbf{l}^\star_1)/(L_c \Pgo)$.
%\end{theorem}

\subsection{Proof of Theorem~\ref{thm:Eqt}}\label{sec:proof_Eqt}

Before proceeding, is it useful to introduce a lemma, which is also leveraged in the proof of Theorem~\ref{thm:Eql}.

%\begin{lemma}%\label{lem:conv}
%	Consider constant prices $\pbar \in \Q^2$ such that $|\pbar_1/\pbar_2 + \wstar_2/\wstar_1|< \delta$ for some $\delta>0$. For sufficiently small $\delta$, there exists an initial AC level distribution for which the sequence $\colK$ converges in distribution to an unique distribution and $\lim_{t\to \infty} \wAt = \lim_{t\to \infty} \nAt=  \nbar$, whereby $\mathbf{1}^\top\nbar = \Pgo$ and  $\pbar^\top \nbar = 0$. Furthermore, $\colL$ converges in probability to a constant $\nbar^\top\mathbf{l}(\nbar)/\Pgo$.
%\end{lemma}
\begin{lemma}\label{lem:conv}
	Consider weight-independent prices $\pbar \in \Q^2$. Define $\nbar \in \Rnn^2$ such that $\mathbf{1}^\top \nbar = \Pgo$ and $\pbar^\top \nbar = 0$. If $l_1(\nbar_1)  < l_2(\nbar_2)$, then: (i)~there exists an initial AC level distribution for which the sequence $\colK$ converges in distribution to an unique AC distribution; (ii)~$\lim_{t\to \infty} \wAt = \lim_{t\to \infty} \nAt=  \nbar$; and (iii)~$\colL$ converges in probability to a constant $\nbar^\top\mathbf{l}(\nbar)/\Pgo$.
\end{lemma}
\begin{proof}
	Consider an initial AC level r.v.\ $K_0$, on which we impose several assumptions. First, we assume that $K_0$ and $W$ are independent, which implies that $K_t$ and $W$ are also independent for any $t\in \N$, since the prices are weight-independent. Second, we assume that, for any  $t \in \Nz$, any $A_t \in \AtNE$ abides by $l_1(\wAt_1) < l_2(\wAt_2)$. Recall that, by Theorem~\ref{thm:NE}, under this condition any $A_t \in \AtNE$ is a Lebesgue-measurable function that is unique up to a set of probability 0. Third, we assume that $P(\{i\in \Omega: K_{0}(i) \in \Kcal\}) = 1$, where $\Kcal := [0, (\Pgo T+1)\pbar_1 - \pbar_2)$. This condition could be relaxed just to a condition on the support of $K_{0}$ being bounded together with the second assumption, since, from Lemma~\ref{lem:brs}, $P(\{i\in \Omega:  \At(i) = 2 \land K_t(i) \geq (\Pgo T+1)\pbar_1 \} = 0$, i.e., the probability of a player with AC level greater or equal to $(\Pgo T+1)\pbar_1$ choosing $\{2\}$ is null. Thus, it follows that there exists a finite $t_1 \in \mathbb{N}$ such that $P(\{i\in \Omega: K_{t_1}(i) \in \Kcal\}) = 1$. Furthermore, by the same argument, the third assumption implies that $P(\{i\in \Omega: K_{t}(i) \in \Kcal\}) = 1,\;\forall t\in \Nz$. In what follows, we show that indeed there exists a $K_0$ for which these conditions hold true, which is trivial for the first and third, but not the second. The components of $\pbar$ are, by hypothesis, rational and nonzero so one can write $\mathbf{p}_1 = a_1/b_1$ and $\mathbf{p}_2 = -a_2/b_2$, where $a_1,b_1,a_2,b_2 \in \mathbb{N}$. Define $\Delta := \gcd(a_1b_2,a_2b_1)/(b_1b_2)$, and notice that the AC level of an individual player with initial AC level $k_0\in \Kcal$ and that chooses strategy $\{1\}$ $x\in \Nz$ times and $\{2\}$ $y\in \Nz$ times can be written as 
	\begin{equation}\label{eq:decomp_AC_level}
		\begin{split}
			\!\!\! k = &(k_0-\lfloor k_0/\Delta\rfloor \Delta) + \\
			&\!\!\!\Delta\left(\!\!\lfloor k_0/\Delta\rfloor \!+ x \frac{a_1b_2}{\gcd(a_1b_2,a_2b_1)} + y\frac{a_2b_1}{\gcd(a_1b_2,a_2b_1)}\!\right)\!\!.\!\!
		\end{split}
	\end{equation}
	Consider a partition of $\Kcal$ denoted by the collection of sets $\{\Kcal_m\}_{m = 1,\ldots,M}$, where $M = \lceil((\Pgo T+1)\pbar_1 - \pbar_2)/\Delta \rceil$, $\Kcal_m = [(m-1)\Delta, m\Delta)$ for $m = 1,\ldots, M-1$, and $\Kcal_M = [(M-1)\Delta, (\Pgo T+1)\pbar_1 - \pbar_2)$. First, from the third assumption, given a $k_0 \in \Kcal$, the AC level of the player will almost surely fall in $\Kcal$ and, thus, in one of the sets $\Kcal_i$. Second, notice that, from \eqref{eq:decomp_AC_level}, given a $k_0 \in \Kcal$, the AC level of the player will always fall in the same relative position inside one of the sets $\Kcal_i$, since $a_1b_2/\gcd(a_1b_2,a_2b_1)$ and $a_2b_1/\gcd(a_1b_2,a_2b_1)$ are integer. Notice that if $k_0-\lfloor k_0/\Delta\rfloor \Delta \geq (\Pgo T+1)\pbar_1 - \pbar_2 - \lfloor((\Pgo T+1)\pbar_1 - \pbar_2)/\Delta \rfloor \Delta$, then $P(\{i\in \Omega: K_{t}(i) \in \Kcal_M\}) = 0, \; \forall t\in \Nz$. In what follows, for simplicity, we just consider the case for which that does not happen, since the analysis in what follows is identical just by disregarding $\Kcal_M$.  Third, from Lemma~\ref{lem:brs}, the second assumption on $K_0$, and since $a_1b_2/\gcd(a_1b_2,a_2b_1)$ and $a_2b_1/\gcd(a_1b_2,a_2b_1)$ are coprime, it follows that there is a path of strategy decisions with nonnull probability such that any set $\Kcal_m$ can be reached from any other set $\Kcal_n$. Therefore, for a given $k_0$ of an individual player, the AC level dynamics can be represented by a Markov chain, whose states are the sets $\{\Kcal_m\}_{m = 1,\ldots,M}$ and whose transition probabilities are time-invariant and given by Lemma~\ref{lem:brs}. Furthermore, this Markov chain has a finite state-space, is irreducible, and is aperiodic (i.e., the transition matrix is primitive and the Markov chain is ergodic). Therefore, it admits an unique stationary distribution~\cite[Proposition 8.4.10]{Rosenthal2006}, which, applying the Perron-Frobenius Theorem~\cite[Theorem~2.12]{Bullo2018} is asymptotically stable. Very similar arguments are employed with more detail in \cite[Theorem~5.1]{SalazarPaccagnanEtAl2021}. Therefore, defining the r.v.s $\{Y_t^m\}_{t \in \Nz, m\in \{1,\ldots,M\}}$ such that $Y_t^m := P(K_t\in \Kcal_m | K_0)$, then it follows that for any $m\in \{1,\ldots,M\}$, $Y_t^m$ converges almost surely to a r.v.\ $Y^m$ (i.e., $P(\lim_{t \to \infty} Y_t^m = Y^m) = 1$), and the convergence is asymptotic. As a result, since, given an initial AC level $k_0$, the player's AC level remains in the same relative position in each $\Kcal_m$, for any $k\in \R$ and any $t\in \Nz$, the r.v.\ $P(K_t\leq k|K_0)$ is uniquely characterized by a sum of $\{Y_t^m\}_{m\in \{1,\ldots,M\}}$ with consecutive superscripts. Therefore, $P(K_t\leq k|K_0)$ also converges almost surely to a unique r.v.\ and the convergence is asymptotic. The c.d.f. of $K_t$, denoted by $F_{K_t}$, is given by $F_{K_t} = \EV[P(K_t\leq k|K_0)]$. Since $P(K_t\leq k|K_0)$ is uniformly bounded by $1$ and converges almost surely to a r.v.\ $P(K\leq k|K_0)$ then, by the Dominated Convergence Theorem \cite[Theorem~9.1.2]{Rosenthal2006},  $\lim_{t\to \infty} F_{K_t}(k) = \EV[P(K\leq k|K_0)] =: F_K(k)$. It follows that $K_t$ converges in distribution. Notice that, since the convergence of the distributions is asymptotic, if the limit distribution abides by the second assumption in $K_0$, then it follows there exists a $K_0$ close enough to the limit distribution such that the whole future trajectory abides by the second assumption. Thus, it is now only necessary to characterize the limit AC level distribution. To that end, one can write $0 = \lim_{t\to\infty} \EV[K_{t}] -\EV[K_{t+1}] = \lim_{t\to\infty} \EV[\pi(W,\At)] = \lim_{t\to\infty} P(A_t =1)\pbar_1 + P(A_t =2)\pbar_2 =  \lim_{t\to\infty} \pbar^\top\nAt$. Since $K_t$ converges in distribution, so does $A_t$ by Theorem~\ref{thm:NE}, thus $\nAt$ converges to some $\nbar \in \Rnn$ such that $\mathbf{1}^\top\nbar = \Pgo$ and  $\pbar^\top\nbar = 0$. By the first assumption on $K_0$, since $K_t$ and $W$ are independent, then $\wAt$ converges to $\nbar$. By hypothesis $l_1(\nbar_1) < \l_2(\nbar_2)$, therefore one concludes that the limit distribution abides by the second assumption in $K_0$, thus there exists a $K_0$ close enough to the limit distribution such that the all assumptions in the beginning of the proof are satisfied. To conclude the proof, it only remains to analyze the limit behavior of $L_t$. To that end, denote the r.v.\ of the number of times that a player chooses strategy $\{1\}$ until time $t$ by $\Nto$. First, we prove that $\Nto/N_t$ converges in probability to $\nbar_1/\Pgo$. This is equivalent to proving that $\pbar_1 \Nto/N_t + (1-\Nto/N_t)\pbar_2$ converges in probability to zero, which follows from  the fact that, for any $\epsilon > 0$ and some $\eta>0$, one can write
	\begin{equation*}
		\begin{split}
			&\lim_{t\to \infty} P(|\pbar_1 \Nto/N_t + (1-\Nto/N_t)\pbar_2| > \epsilon)\\
			=& \lim_{t\to \infty} P((K_t-K_0)/N_t <- \epsilon)+ P((K_t-K_0)/N_t > \epsilon)\\
			\leq &  \lim_{t\to \infty} P(K_0\!-\!\epsilon N_t \!\geq\! 0) \!+\!P(\epsilon N_t \!+\!K_0 \!<\! (\Pgo T\!+\!1)\pbar_1 \!-\! \pbar_2)\\
			\leq & \lim_{t\to \infty} 2P(N_t/t \leq ((\Pgo T\!+\!1)\pbar_1 \!-\! \pbar_2)/(t\epsilon))\\
			\leq & \lim_{t\to \infty} 2P(N_t/t - \Pgo \leq -\eta)\\
			\leq & \lim_{t\to \infty} 2P(|N_t/t - \Pgo| > \eta) =  0,
		\end{split}
	\end{equation*}
	where the first inequality follows from the fact that $P(K_t \notin \Kcal) = 0 \; \forall t\in \Nz$, the second inequality follows from $ \max(K_0/(t\epsilon),((\Pgo T\!+\!1)\pbar_1 \!-\! \pbar_2-K_0)/(t\epsilon)) \leq ((\Pgo T\!+\!1)\pbar_1 \!-\! \pbar_2)/(t\epsilon)$, the third inequality follows from particularizing the RHS of the inequality inside the probability with $t = ((\Pgo T\!+\!1)\pbar_1 \!-\! \pbar_2)/(\epsilon (\Pgo -\eta)) < \infty$ for some $\eta >0$, and the last equality follows from the weak law of large numbers \cite[Theorem~5.3.1]{Rosenthal2006} on $\colP$. It follows immediately that, since $L_t$ can be written as $L_t = l_1(\wAt_1) \Nto/N_t + (1-\Nto/N_t)l_2(\wAt_2)$,  $L_t$ converges in probability to  $\nbar^\top\mathbf{l}(\nbar)/\Pgo$.
\end{proof}

%\begin{theorem}%\label{thm:Eqt}
%	For a sufficiently small $\epsilon >0$, consider a pricing policy
%	\begin{equation*}
%		\mathbf{p}(w) = {S[\rat_\delta(\wstar_2/\wstar_1)\, -\!1]^\top},
%	\end{equation*}
%	where $\delta = (\epsilon\Pgo C(\wstar)/(\wstar_1(L_C\wstar_1 +\epsilon C(\wstar))$ and $S \in \Q_{>0}$. Then, there exists an initial AC level distribution for which $\PoA_t$ and $\InEqt_t$ converge and $\lim_{t \to \infty} \PoA_t \leq 1+ \epsilon$ and $\lim_{t \to \infty} \InEqt_t = 0$. %Furthermore, if $\EV[1/W^2]\leq \infty$, then $\InEql_t$ converges and $|\lim_{t \to \infty} \InEql_t - ({\mathbf{l}^\star}^\top \mathbf{w}^\star/\Pgo)(\EV[1/W^2]-\EV[1/W]^2)| \leq \epsilon(\mathbf{l}^\star_2 -\mathbf{l}^\star_1)/(L_c \Pgo)$.
%\end{theorem}

Consider $\pbar = S[\rat_\delta(\wstar_2/\wstar_1)\, -\!1]^\top$ for a sufficiently small $\delta$. Define $\nbar \in \Rnn^2$ such that $\mathbf{1}^\top \nbar = \Pgo$ and $\pbar^\top \nbar = 0$. By the definition of the rational approximation function $\rat_\delta$, $|\wstar_2/\wstar_1 - \nbar_2/\nbar_1|<\delta$, therefore, for sufficiently small $\delta$, since $\mathbf{l}$ is continuous, it follows that $l_1(\nbar_1) < \l_2(\nbar_2)$. Therefore, one may apply Lemma~\ref{lem:conv}. It follows that there exists $K_0$ such that $\lim_{t\to \infty} \nAt=  \nbar$. As a result, $|\nbar_1 - \wstar_1| \leq \delta \wstar_1/(\Pgo/\wstar_1-\delta)$, and by the assumption that $C$ is Lipschitz continuous, one concludes that $\lim \PoA_t \leq 1+ \epsilon$ with $\epsilon =   (L_C/(C(\wstar))\delta \wstar_1/(\Pgo/\wstar_1-\delta)$. Furthermore, since, by Lemma~\ref{lem:conv}, $\colL$ converges in probability to $\nbar^\top\mathbf{l}(\nbar)/\Pgo$ and $\colL$ is uniformly bounded, it follows by the Dominated Convergence Theorem~\cite[Theorem~9.1.2]{Rosenthal2006} that $\lim_{t\to \infty} \EV[L_t] = \nbar^\top\mathbf{l}(\nbar)/\Pgo$ and $\lim_{t \to \infty} \InEqt_t = \lim_{t \to \infty} \sqrt{\Var[L_t]} = 0$. $\hfill\square$  %Since the pricing policy does not depend on the weight of the players, $L_t$ and $W$ are independent and, if $\EV[1/W^2]<\infty$, then  $\lim_{t \to \infty} \InEql_t = \lim_{t \to \infty} \sqrt{\Var[L_t/W]} = \EV[L_t]\sqrt{(\EV[1/W^2]-\EV[1/W]^2)}$.

\subsection{Proof of Lemma~\ref{lem:sol_theta}}\label{sec:proof_sol_theta}

Define $\underline{\theta}:= \wmin(\xi -\wstar_1)/\xi$ and $\bar{\theta}:= \wmax(\xi-\wstar_1)/(\xi-\Pgo)$.  Define $\tilde{n}_1:\R\times [\underline{\theta},\bar{\theta}] \to \R$ as a domain extension of $n_1$ such that $\tilde{n}_1(w,\theta) = n_1(w,\theta)$, if $w \in [\wmin,\wmax]$, and $\tilde{n}_1(w,\theta) = 0$, otherwise, for all $\theta \in[\underline{\theta},\bar{\theta}]$. 
Define $H:[\underline{\theta},\bar{\theta}] \to \R$ as $H(\theta) := \wstar_1\EV[W] - \int_\Omega \tilde{n}_1(W,\theta)W \dint P$.  
Since $\tilde{n}_1(\cdot,\theta)$ is piecewise-continuous for any $\theta \in[\underline{\theta},\bar{\theta}]$, it is measurable and, thus, $H$ is well-defined. To prove the continuity of $H$ in  $[\underline{\theta},\bar{\theta}]$ consider a sequence $\theta_n$ in $[\underline{\theta},\bar{\theta}]$ such that $\lim \theta_n = \theta \in [\underline{\theta},\bar{\theta}]$. Since $\tilde{n}_1(w,\cdot)$ is continuous for all $w\in \R$, then $\lim W\tilde{n}_1(W,\theta_n) = W\tilde{n}_1(W,\theta)$. Furthermore, since $|\tilde{n}_1(w,\theta)| \leq \Pgo \mathbf{1}_{\Wcal}(w)$ for all $w\in \R$ and $\theta \in [\underline{\theta},\bar{\theta}]$ and $\mathbf{1}_{\Wcal}(w)$ is Lebesgue-integrable, it follows by the Dominated Convergence Theorem \cite[Theorem~9.1.2]{Rosenthal2006} that $\lim H(\theta_n) = H(\theta)$. Therefore $H$ is continuous in $[\underline{\theta},\bar{\theta}]$. Furthermore, notice that $n_1(\cdot,\underline{\theta}) \equiv 0$ and $n_1(\cdot,\bar{\theta}) \equiv \Pgo$, thus $H(\underline{\theta}) = \wstar_1\EV[W] >0$ and  $H(\bar{\theta}) = -(\Pgo-\wstar_1)\EV[W] <0$. It follows by the Intermediate Value Theorem that there is $\theta^\star \in [\underline{\theta},\bar{\theta}]$ such that $H(\theta^\star) = 0$, thus \eqref{eq:design_theta} admits at least one solution. 
Consider two solutions $\theta^\star_1,\theta^\star_2 \in [\underline{\theta},\bar{\theta}]$ to \eqref{eq:design_theta} and admit, without any loss of generality, that $\theta^\star_1 \leq \theta^\star_2$. Since they are solutions to \eqref{eq:design_theta}, then
\begin{equation*}
	\int_\Omega n_1(W,\theta^\star_1)W \dint P = 	\int_\Omega n_1(W,\theta^\star_2)W \dint P.
\end{equation*}
Furthermore, notice that  $n_1(w,\theta^\star_1) =  n_1(w,\theta^\star_2)$ for all $w\in \Wcal_{1,2}$, with $\Wcal_{1,2}:=[\wmin, \theta_1^\star(1-(\Pgo-\wstar_1)/(\xi-\wstar_1))] \cup [\theta_2^\star(1+\wstar_1/(\xi-\wstar_1)), \wmax]$, and $wn_1(w,\theta^\star_1) < wn_2(w,\theta^\star_2)$ for all $w\in \Wcal\setminus \Wcal_{1,2}$. Therefore, it must be the case that $P(W\in \Wcal\setminus \Wcal_{1,2}) = 0$. Consider the average perceived latency r.v. prescribed by $\theta_1^\star$ and $\theta_2^\star$, denoted by $L_1$ and $L_2$, respectively, which are given by $L_i = \left(n_1(W,\theta^\star_i)\lstar_1 + (\Pgo-n_1(W,\theta^\star_i)) \lstar_2\right)/\Pgo$ with $i\in\{1,2\}$. Since, $wn_1(w,\theta^\star_1) =  wn_1(w,\theta^\star_2)$ for all $w\in \Wcal_{1,2}$ and $P(W\in \Wcal\setminus \Wcal_{1,2}) = 0$, it follows that $\EV[L_1] = \EV[L_2]$ and $\Var[L_1] = \Var[L_2]$. Thus, the inequality prescribed by any two solutions $\theta_1^\star$ and $\theta_2^\star$ is the same, i.e., $\sqrt{\Var[L_1]} = \sqrt{\Var[L_2]}$. $\hfill\square$

\subsection{Proof of Theorem~\ref{thm:Eql}}\label{sec:proof_Eql}

%The pricing policy can be rewritten as 
%	\begin{equation*}\small
%	\mathbf{p}(w) \!=\!\! 
%	\begin{cases}
%		\!S\left[0\ -\!1\right]^\top\!\!, & \frac{w}{\theta^\star} -1 \leq -\frac{\Pgo-\wstar_1}{\xi-\wstar_1}\\
%		\!S\!\left[\rat_\delta\!\left(\frac{n_2(w,\theta^\star)}{n_1(w,\theta^\star)}\right)\; -\!1\right]^\top\!\!\!, \!\!&\!\!\!\!  \!\!-\frac{\Pgo-\wstar_1}{\xi-\wstar_1}\! <\! \frac{w}{\theta^\star} \!-\!1 \!\leq \!0\\
%		\!S\rat_\delta\!\left(\frac{\wstar_2}{\wstar_1}\right) \begin{bmatrix}1 \\\!-\!\rat_\delta\!\left(\frac{n_2(w,\theta^\star)}{n_1(w,\theta^\star)}\right)\end{bmatrix}\!\!,\!\! & \!\!0 <\frac{w}{\theta^\star}\! -\! 1 \!<\! \frac{\wstar_1}{\xi-\wstar_1},\\
%		\!S\rat_\delta\!\left(\frac{\wstar_2}{\wstar_1}\right) \left[1 \;0\right]^\top\!\!, & \frac{w}{\theta^\star} - 1 \geq \frac{\wstar_1}{\xi-\wstar_1}.
%	\end{cases}
%\end{equation*}
Notice that, from the definition of the rational approximation function $\rat_\delta$, the pricing policy takes a finite number of prices. Recall that there are intervals $\Acal_j := ((j-1)\delta,j\delta]$ for which $\rat_\delta(x) := \sum_{j=1}^\infty q_j\mathbf{1}_{\Acal_j}(x)$, where $q_j \in \Qp$ satisfies $|q_j-x|<\delta\; \forall x\in \Acal_j$. Indeed, one may partition $\Wcal$ in a finite number of weight brackets $\{\Wcal_j\}_{j = 0,\ldots,M_1+M_2+1}$ for which the prices are weight-independent. Indeed, these are defined by $\Wcal_0:= \{w\in \Wcal : n_1(w,\theta^\star) = \Pgo\}$, 
\begin{equation*}
	\Wcal_j := \left \{\!w\!\in\! \Wcal : \frac{n_2(w,\theta^\star)}{n_1(w,\theta^\star)} \in \Acal_j\!\! \right\}\!, \, j \!= \!1,\ldots,M_1,
\end{equation*}
where $M_1 := \lceil \wstar_2/(\wstar_1 \delta)\rceil$, 
\begin{equation*}
	\Wcal_j \!:= \!\left \{\!w\!\in\! \Wcal : \frac{n_1(w,\theta^\star)}{n_2(w,\theta^\star)} \in \Acal_{j-M_1}\!\!\right\}\!, \, j \!= \!1\!+\!M_1,\ldots,M_2,
\end{equation*}
where $M_2 := \lceil \wstar_1/(\wstar_2 \delta)\rceil$, and $\Wcal_{M_1+M_2+1}:= \{w\in \Wcal : n_1(w,\theta^\star) = 0\}$. Similarly, we may also partition $\Omega$ analogously in a finite number of sets $\{\Omega_j\}_{j = 0,\ldots,M_1+M_2+1}$, defined by $\Omega_j := \{i\!\in\! \Omega: W(i)\!\in\! \Wcal_j\}$. Define $\pbarOj$ as the weight-independent price vector that the players in $\Omega_j$ are subject to.  Define $\nbarOj \in \Rnn^2$ such that $\mathbf{1}^\top \nbarOj = \Pgo$ and $\pbarOj\!\!\phantom{l}^\top \nbarOj = 0$. Similarly to Lemma~\ref{lem:conv}, the goal is now to prove that there exists an initial AC level r.v.\ $K_0$ for which $P(A_t = 1|W\in \Wcal_j)$ converges, for each $\Omega^j$, to $\nbarOj$.  First, since $\pbar^{\Omega_0} = S [0\; -1]^\top$, strategy $\{1\}$ is always chosen by the players in $\Omega_0$ for any $t\in \Nz$, thus $P(A_t = 1|W\in \Wcal_0) = \Pgo = \nbar^{\Omega_0}_1$ for any $K_0$.  %\nbar^{\Omega_0} = \Pgo [1 \;0]^\top
Second, since $\pbar^{\Omega_{M_1+M_2+1}} = S \rat_\delta(\wstar_2/\wstar_1)[1\; 0]^\top$, strategy $\{2\}$ is always chosen by the players in $\Omega_{M_1+M_2+1}$ if their AC level is initialized to be below $S \rat_\delta(\wstar_2/\wstar_1)$. As a result, for any $t\in \Nz$ it follows that $P(A_t = 1|W\in \Wcal_{M_1+M_2+1}) = 0 = \nbar^{\Omega_{M_1+M_2+1}}_{1}$. Finally, for $j \in \{1,\ldots,M_1+M_2\}$, following the exact same steps of the proof of Lemma~\ref{lem:conv} for the subpopulation $\Omega_j$, which are omitted for the sake of conciseness, it follows that there exists $K_0$ such that $P(A_t = 1|W\in \Wcal_{j})$ converges to $\nbarOj_1$, if the limit of the aggregate of all subpopulations is in the region whereby $l_1(\wAt_1)  < l_2(\wAt_2)$. Therefore, one must characterize the aggregate of the limit. On the one hand, denote the proportion of the cumulative weight of the players at the limit by $\wbar \in \Rnn$, which follows $\mathbf{1}^\top\wbar = \Pgo$ and
\begin{equation}\label{eq:wr_limit}
	\begin{split}
		\wbar_1  &= \int_\Omega P(\At = r|W)W\dint P /\EV[W]\\
			& = \int_\Omega \sum\nolimits_{j = 0}^{M1+M_2+1}\mathbf{1}_{\Wcal_j}(W)\nbarOj_1W\dint P /\EV[W].
	\end{split}
\end{equation} 
On the other hand, it follows from the fact that $\theta^\star$ is a solution to \eqref{eq:design_theta} that 
\begin{equation}\label{eq:wstar_limit}
		\wstar_1 = \int_\Omega n_1(W,\theta^\star)W \dint P/ \EV[W].
\end{equation}
To relate \eqref{eq:wr_limit} and \eqref{eq:wstar_limit}, notice that: (i)~$|\nbar^{\Omega_0}_1-n_1(w,\theta^\star)| = 0,\; \forall w\in \Wcal_0$; (ii)~$|\nbarOj_1-n_1(w,\theta^\star)| < \delta/\Pgo, \, \forall j \in \{1,\ldots,M_1\}\, \forall w\in \Wcal_{j}$; (iii)~$|\nbarOj_1-n_1(w,\theta^\star)| < \delta\Pgo, \, \forall j \in \{M_1+1,\ldots,M_1+M_2\}\, \forall w\in \Wcal_{j}$; and (iv)~$|\nbar^{\Omega_{M_1+M_2+1}}_1-n_1(w,\theta^\star)| = 0,\; \forall w\in \Wcal_{M_1+M_2+1}$.
The inequalities above follow, after algebraic manipulations, from the definition of $\rat_\delta$ and $\pbarOj\!\!\phantom{l}^\top \nbarOj = 0$. Since $\delta/\Pgo > \delta\Pgo$, then one may conclude that $|\nbarOj_1-n_1(w,\theta^\star)| < \delta/\Pgo, \, \forall j \in \{0,\ldots,M_1+M_2+1\}\, \forall w\in \Wcal_{j}$. As a result, from  \eqref{eq:wr_limit} and \eqref{eq:wstar_limit}, 
\begin{equation*}
	\begin{split}
	|\wbar_1\!-\!\wstar_1| & \!\leq \!\int_\Omega \sum_{j = 0}^{ \!M_1\!+\!M_2\!+\!1}\!\!\!\mathbf{1}_{\Wcal_j}(W)\!\left|\nbarOj_1\!-\!n_1(W,\theta^\star)\right|\!\frac{W}{\EV[W]}\dint P\\	
	& <	\delta/\Pgo.
\end{split}
\end{equation*}
Therefore, for sufficiently small $\delta$, since $\mathbf{l}$ is continuous, it follows that $l_1(\wbar_1) < \l_2(\wbar_2)$. Therefore, there exists $K_0$ such that $\lim_{t\to \infty} P(\At = 1|W\in \Wcal_j) = \nbarOj_1, \, \forall j\in \{0,\ldots,M_1\!+\!M_2\!+\!1\}$. As a result, by the assumption that $C$ is Lipschitz continuous, one concludes that $\lim \PoA_t \leq 1+ \epsilon$ with $\delta =  \epsilon\Pgo C(\wstar)/L_C$. Turning now to the convergence of $\InEql_t$, on the one hand, by the same arguments employed in Lemma~\ref{lem:conv}, but now applied to each subpopulation $\Omega_j$, it follows that for any $\eta>0$
\begin{equation*}
	\lim_{t \to \infty} P\left(\left|L_t- \frac{ \Pgo\lstar_2 -\nbarOj_1(\lstar_2-\lstar_1)}{\Pgo} \right| > \eta \,\Bigg|\, W \!\in\! \Wcal_j\right) = 0, 
\end{equation*}
thus $L_t$ converges in probability to 
\begin{equation*}
	L = \frac{1}{\Pgo}\left( \Pgo \lstar_2 -  \sum_{j = 0}^{ \!M_1+M_2+1}\!\!\!\mathbf{1}_{\Omega_j}\nbarOj_1(\lstar_2-\lstar_1)\right)
\end{equation*}
and, since $L_t$ and $L_t/W$ are uniformly bounded, then $\EV[L_t]$ and  $\EV[L_t/W]$ converge to $\EV[L]$ and $\EV[L/W]$, respectively, by the Dominated Convergence Theorem~\cite[Theorem~9.1.2]{Rosenthal2006}.
Finally, the average perceived latency r.v. prescribed by $n_1(W,\theta^\star)$ is  given by 
\begin{equation*}
	L^\star= \left(\Pgo\lstar_2 -n_1(W,\theta^\star_i) (\lstar_2-\lstar_1)\right)/\Pgo,
\end{equation*}
which achieves optimal inequality $\InEql^\star$. After algebraic manipulations, it follows that: (i)~$|\EV[L/W]-\EV[L^\star/W] < \delta(\lstar_2-\lstar_1)\EV[1/W]/\Pgo^2$; (ii)~$|\EV[L/W]^2-\EV[L^\star/W]^2| < D_1\delta$; and (iii)~$|\EV[(L/W)^2]-\EV[(L^\star/W)^2]| < D_2\delta$, where $D_1:=2\lstar_2(\lstar_2-\lstar_1)\EV[1/W]/(\Pgo^3\wmin)$ and $D_2 := 2(\lstar_2-\lstar_1)(2\lstar_2-\lstar_1)\EV[1/W^2]/\Pgo^2$. Therefore, $|\lim_{t \to \infty}\InEql_t^2-{\InEql^\star}^2| < D\epsilon$ and $|\lim_{t \to \infty}\InEql_t-{\InEql^\star}|< \sqrt{D\epsilon}$, where $D:= (D_1+D_2)\Pgo C(\wstar)/ L_C$. $\hfill\square$

\fi

\bibliographystyle{IEEEtran}
%\bibliography{../../_bib/references-gt.bib}
\bibliography{../../../Bibliography/references-gt.bib,../../../Bibliography/main.bib,../../../Bibliography/SML_papers.bib}
%\bibliography{references-gt.bib}

%\theendnotes

\end{document}